\documentclass[aps,twocolumn,superscriptaddress,amsmath,amssymb,prl]{revtex4-2}
\usepackage{graphicx}
\usepackage{dcolumn}
\usepackage{bm}
\usepackage{braket}
\usepackage{appendix}
\usepackage{subfigure}
\usepackage{hyperref}
\usepackage{color}
\usepackage{xcolor}
\usepackage{amsmath}
\usepackage{nccmath}
\usepackage[english]{babel}
\usepackage{svg}
\begin{document}

\title[High-contrast shallow spin defects]
 {High-contrast plasmonic-enhanced shallow spin defects in hexagonal boron nitride for quantum sensing}
 
\author{Xingyu Gao}
\affiliation{Department of Physics and Astronomy, Purdue University, West Lafayette, Indiana 47907, USA}
\author{Boyang Jiang}
\affiliation{School of Electrical and Computer Engineering, Purdue University, West Lafayette, Indiana 47907, USA}
\author{Andres E. Llacsahuanga Allcca}
\affiliation{Department of Physics and Astronomy, Purdue University, West Lafayette, Indiana 47907, USA}
\author{Kunhong Shen}
\affiliation{Department of Physics and Astronomy, Purdue University, West Lafayette, Indiana 47907, USA}
\author{Mohammad A. Sadi}
\affiliation{School of Electrical and Computer Engineering, Purdue University, West Lafayette, Indiana 47907, USA}
\author{Abhishek B. Solanki}
\affiliation{School of Electrical and Computer Engineering, Purdue University, West Lafayette, Indiana 47907, USA}
\author{Peng Ju}
\affiliation{Department of Physics and Astronomy, Purdue University, West Lafayette, Indiana 47907, USA}
\author{Zhujing Xu}
\affiliation{Department of Physics and Astronomy, Purdue University, West Lafayette, Indiana 47907, USA}
\author{Pramey Upadhyaya}
\affiliation{School of Electrical and Computer Engineering, Purdue University, West Lafayette, Indiana 47907, USA}
\author{Yong P. Chen}
\affiliation{Department of Physics and Astronomy, Purdue University, West Lafayette, Indiana 47907, USA}
\affiliation{School of Electrical and Computer Engineering, Purdue University, West Lafayette, Indiana 47907, USA}
\affiliation
{Birck Nanotechnology Center, Purdue University, West Lafayette,
	IN 47907, USA}
\affiliation
{Purdue Quantum Science and Engineering Institute, Purdue University, West Lafayette, Indiana 47907, USA}
\author{Sunil A. Bhave}
\affiliation{School of Electrical and Computer Engineering, Purdue University, West Lafayette, Indiana 47907, USA}
\affiliation
{Birck Nanotechnology Center, Purdue University, West Lafayette,
	IN 47907, USA}
\author{Tongcang Li}
\email{tcli@purdue.edu}
\affiliation{Department of Physics and Astronomy, Purdue University, West Lafayette, Indiana 47907, USA}
\affiliation{School of Electrical and Computer Engineering, Purdue University, West Lafayette, Indiana 47907, USA}
\affiliation
{Birck Nanotechnology Center, Purdue University, West Lafayette,
	IN 47907, USA}
\affiliation
{Purdue Quantum Science and Engineering Institute, Purdue University, West Lafayette, Indiana 47907, USA}

\date{\today}



\begin{abstract}
The recently discovered spin defects in hexagonal boron nitride (hBN), a layered van der Waals material, have great potential in quantum sensing. However, the photoluminescence and the contrast of the optically detected magnetic resonance (ODMR) of hBN spin defects are relatively low so far, which limits their sensitivity. Here we report a record-high ODMR contrast of 46$\%$ at room temperature, and simultaneous enhancement of the photoluminescence of hBN spin defects by up to 17-fold by the surface plasmon of a gold-film microwave waveguide. Our results are obtained with shallow boron vacancy spin defects in hBN nanosheets created by low-energy He$^+$ ion implantation, and a gold-film microwave waveguide fabricated by photolithography.  We also explore the effects of  microwave and laser powers on the ODMR, and improve the sensitivity of hBN spin defects for magnetic field detection. Our results support the promising potential of hBN spin defects for nanoscale quantum sensing.
\end{abstract}

\maketitle

\begin{figure}[tbh]
	\centering
	\includegraphics[width=0.49\textwidth]{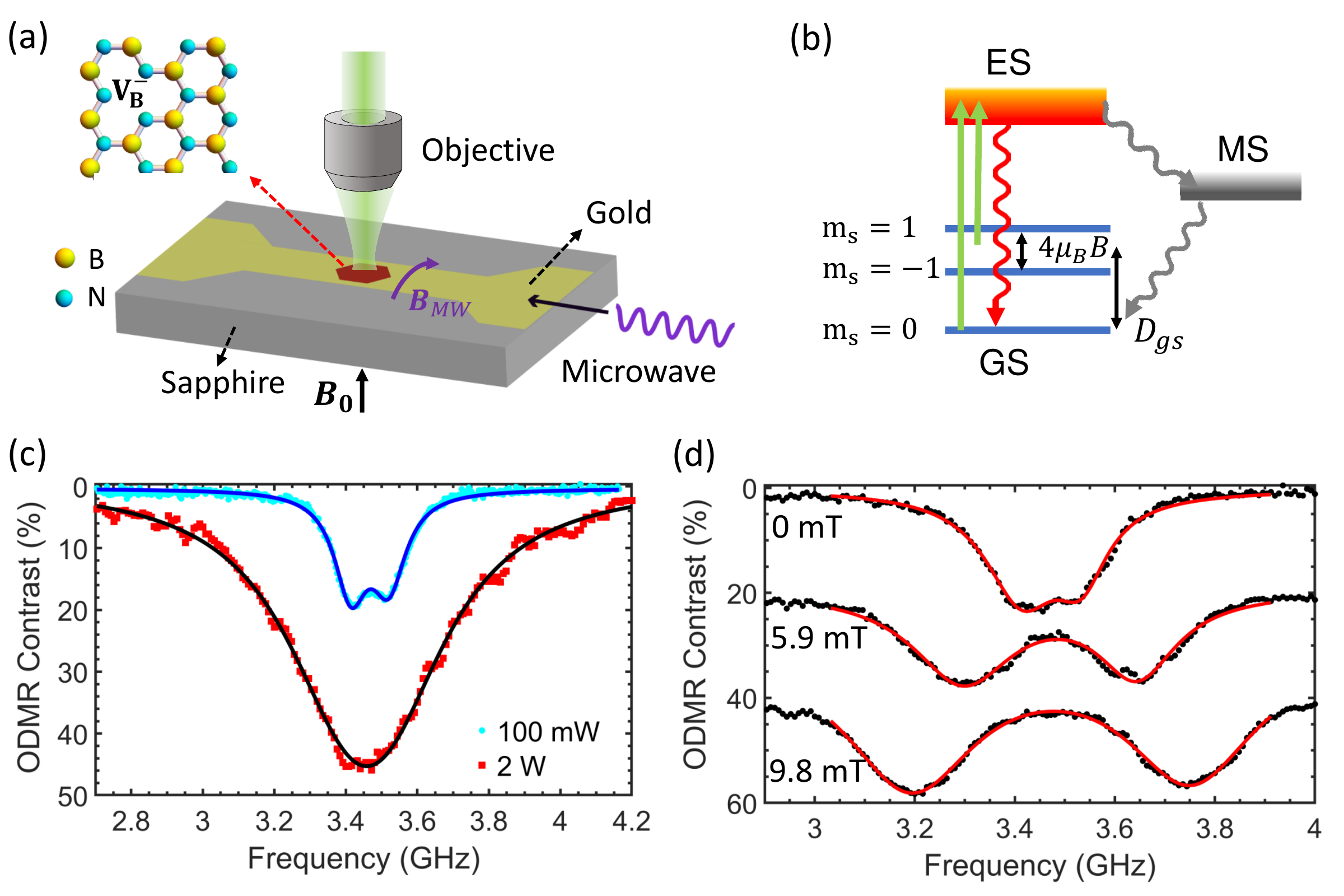}
	\caption{ (a) An illustration of the experimental setup for ODMR measurements. An ion implanted hBN nanosheet is placed on top of a gold-film microwave stripline. A microwave is delivered through the stripline for spin manipulation. The surface plasmon of the gold surface provides emission enhancement. An NA=0.9 objective lens is used to excite $V_B^-$ defects with a 532-nm laser and collect the PL. A permanent magnet is used to apply a static magnetic field. (b) The energy diagram of a $V_B^-$ defect and  the optical pumping cycle between the ground state (GS), the excited state (ES) and the metastable state (MS). A magnetic field  induce Zeeman shifts of the  spin sublevels. (c) Measured CW ODMR spectra under 100 mW (light blue dots) and 2W  (red squares) microwave  driving. With a high-power microwave (2 W), the ODMR contrast can reach 46$\%$. The laser excitation power is 5 mW. The $V_B^-$ defects are generated by 2.5 keV He$^+$ ion implantation. (d) ODMR spectra in  different external magnetic fields. A clear splitting  of 346 MHz and 560 MHz is observed in 5.9 mT and 9.8 mT magnetic fields, respectively. Solid curves are fittings with a double Lorentzian model.} \label{fig:1}
\end{figure}

Optically active spin defects in wide-band-gap materials have shown great potential for  a wide range of emerging technologies, from quantum information processing \cite{togan2010quantum,bradley2019ten} to high-resolution sensing of magnetic and electric fields\cite{dolde2011electric,grinolds2014subnanometre,schirhagl2014nitrogen,Thiel973}. Color centers in bulk semiconductors such as diamond \cite{doherty2013nitrogen,gruber1997scanning} and silicon carbide\cite{koehl2011room,riedel2012resonant} are prime examples that reveal optically detected magnetic resonance (ODMR).
Recently, atomic defects in layered van der Waals materials such as hexagonal boron nitride (hBN) are attracting increasing attention as alternative candidates for studying light-matter interaction, nanophotonics and nanoscale sensing\cite{tran2017deterministic,caldwell2019photonics,exarhos2019magnetic,Konthasinghe19}. 
Atomic defects in hBN are stable in nanosheets as thin as a monolayer and are readily accessible for device integration and top down nanofabrication \cite{tran2016quantum,palombo2020nanoscale}. Furthermore, recent experiments discovered that some defects in hBN could be spin addressable at room temperature\cite{gottscholl2020initialization,mendelson2021identifying,gottscholl2021room,stern2021room,gottscholl2021sub}. The negatively charged boron vacancy ($V_B^-$) spin defect  is the most studied one among these defects \cite{ivady2020ab,Reimers2020PhysRevB.102.144105}. It has spin $s=1$ and its orientation is out of plane \cite{gottscholl2020initialization}. $V_B^-$ defects can be  generated by neutron irradiation \cite{gottscholl2020initialization,liu2021rabi}, ion implantation\cite{kianinia2020generation}, femtosecond laser writing \cite{gao2021femtosecond}, and electron irradiation \cite{nano11061373}. Spin defects in thin hBN nanosheets will be useful for quantum sensing and spin optomechanics \cite{AbdiPRL2017,LiPRL2020spinphonon}.  However, so far the $V_B^-$ spin defects have relatively low brightness and ODMR contrast, which limit their sensitivity  \cite{gottscholl2021sub}.

Here we report high-contrast plasmonic-enhanced  shallow $V_B^-$ spin defects in hBN nanosheets for quantum sensing.  
We fabricate a gold-film coplanar microwave waveguide  by photolithography  to optimize the homogeneity and local intensity of the microwave for spin control. The hBN nanosheets with spin  defects are transferred onto the microwave waveguide and are in contact of the gold surface for  both plasmonic emission enhancement and spin control. The surface plasmons \cite{Shimizu2002PRLsurface,song2014photoluminescence,shalaginov2020chip} provide broadband emission enhancement covering the wide range of the photoluminescence (PL) of $V_B^-$ defects from 750 nm to 950 nm \cite{gottscholl2020initialization}.  Our method does not require complex nanofabrication or cause adverse effects on quantum sensing.
The microwave magnetic field generated by the gold waveguide is parallel to the surface and perpendicular to the orientation of $V_B^-$ electron spins, which is crucial to achieve high ODMR contrast.
With these, we find that the ODMR contrast can reach 46$\%$ at room temperature, which is   an order of magnitude larger than the highest room-temperature ODMR contrast of hBN spin defects reported in the previous work \cite{liu2021temperature}.  We also observe an up to 17-fold photoluminescence (PL) enhancement of $V_B^-$ defects due to the gold film. We measure the spin initialization time to be around 100~ns. In addition, we study the laser power and microwave power dependence of the continuous-wave (CW) ODMR, and optimize the magnetic field sensitivity to be about 8 $\mu \rm T/\sqrt{\rm Hz}$. Finally, we perform coherent spin control and measure the spin-lattice relaxation time  $T_1$ and the spin coherence time $T_2$ of  $V_B^-$ spin defects generated by He$^+$ ion implantation, which can benefit future works on multi-pulse sensing protocols. Our results demonstrate the promising potential of hBN spin defects  for nanoscale quantum sensing and other quantum technologies.

\begin{figure}[tbh]
	\centering
	\includegraphics[width=0.48\textwidth]{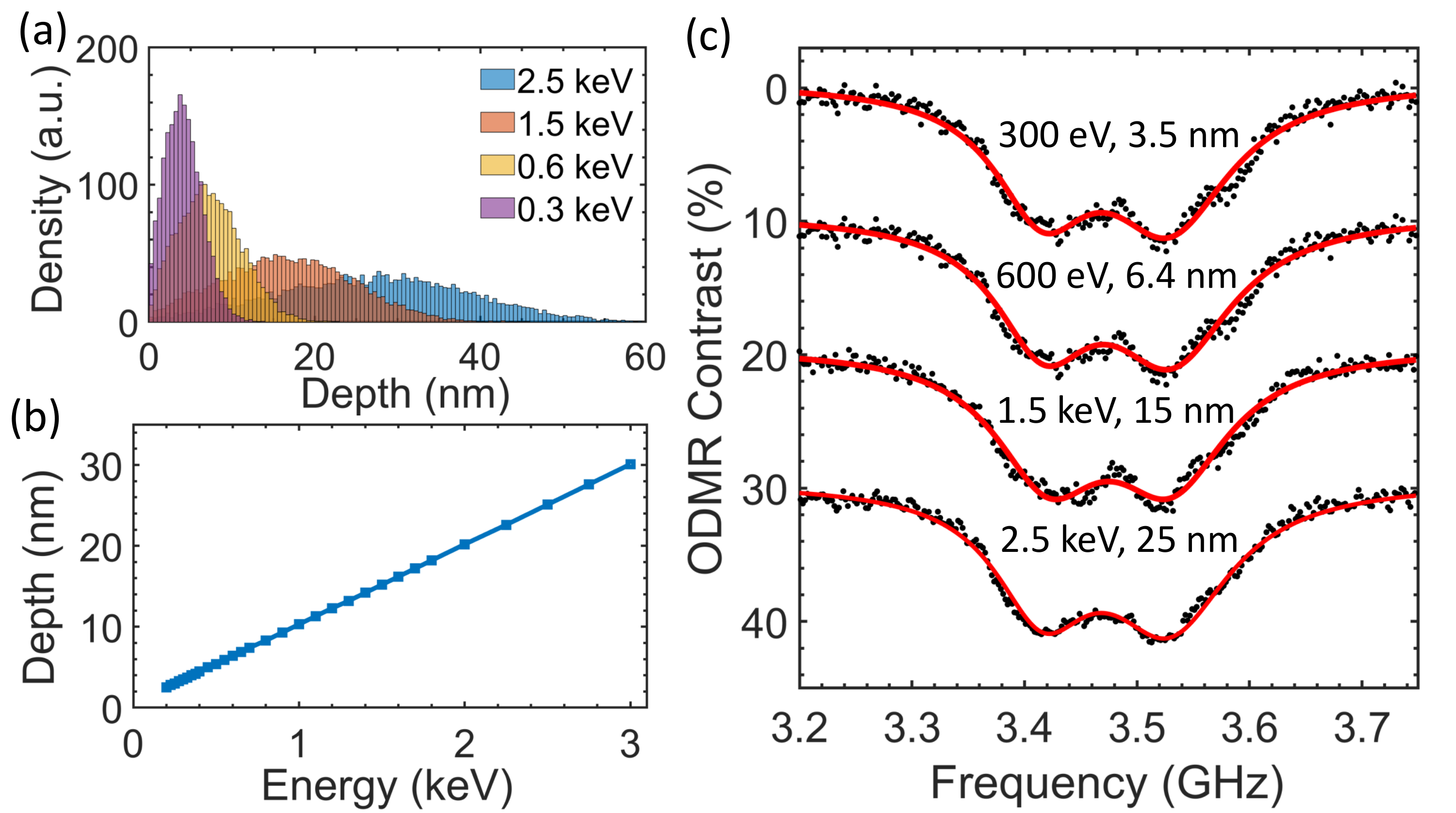}
	\caption{ (a) The depth distribution of defects created by He$^+$ ions with different implantation energies. The results are obtained with SRIM simulation. (b) The most probable depth to create defects as a function of ion energies. (c) Measured ODMR spectra of $V_B^-$ with different implantation depths. The linewidths are 122 MHz, 115 MHz, 104 MHz, 103.3 MHz for 3.5 nm, 6.2 nm, 15 nm and 25 nm depths, respectively.} \label{fig:depth}
\end{figure}

The results presented in this work are obtained with tape-exfoliated hBN nanosheets which are tens of nanometers thick. We exfoliate hBN flakes onto a Si wafer, and then mount the wafer in a home-built ion implanter for doping. We use low-energy He$^+$ ions (200 eV - 3 keV) to implant hBN nanosheets, which creates high-quality $V_B^-$ defects with average depths ranging from 3 nm to 30 nm and avoids introducing  undesired defects \cite{lehtinen2011production}.  After ion implantation, the hBN flakes are transferred onto a gold microwave stripline on a sapphire substrate and characterized using an ODMR setup (Fig. \ref{fig:1} (a)). CW ODMR measurements are performed to acquire the basic spin properties of the $V_B^-$ defects generated by He$^+$ ions. The defects are excited by a 532 nm laser with an NA=0.9 objective lens, which also collects the PL of defects. We record integrated photon counts as a function of the applied microwave. We obtain the difference between the photon count rates when the microwave is off ($I_{off}$) and when the microwave is on ($I_{on}$). The ODMR contrast is then determined by normalizing the PL difference with the PL intensity when the microwave is off: $C=(I_{off}-I_{on})/I_{off}$. A positive ODMR contrast $C$ means that the microwave driving  decreases the PL intensity. 

Figure \ref{fig:1}(c) shows measured high-contrast ODMR spectra of $V_B^-$ defects at room temperature. Strikingly, with  strong microwave driving, we find that these defects can exhibit up to 46$\%$ ODMR contrast, which is one order of magnitude higher than the best contrast of hBN spin defects reported previously \cite{liu2021temperature}. This value is even larger than the ODMR contrast of diamond nitrogen-vacancy centers at room temperature \cite{doherty2013nitrogen}. The ODMR contrast can readily reach around 20$\%$ without significant power broadening using low-power microwave driving.  Without an external magnetic field, the measured ODMR spectrum shows two resonances at $\nu_1$ and $\nu_2$ centered around $\nu_0$ (Fig. \ref{fig:1}(c)). The results agree well with a double Lorentzian model. Here $\nu_0$ is determined by the zero-field splitting (ZFS), $\nu_0=D_{gs}/h$= 3.47 GHz, where $h$ is the Planck constant. And the splitting between $\nu_1$ and $\nu_2$ is due to the non-zero off-axial ZFS parameter $E_{gs}/h=$ 50 MHz \cite{gottscholl2020initialization}. Fig. \ref{fig:1} (d) presents ODMR spectra in different external static magnetic fields. We use a permanent magnet to apply a static magnetic field perpendicular to the nanosheet surface.  A translation stage is used to change the position of the magnet and tune the magnetic field strength. With an external static magnetic field $\mathbf{B}$, $\nu_1$ and $\nu_2$ will be split  further owing to the Zeeman effect, $\nu_{1,2}=D_{gs}/h\pm\sqrt{E_{gs}^2+(g\mu_BB)^2}/h$, where $g=2$ is the Landé $g$-factor. The splitting ($\nu_2-\nu_1$) is 560 MHz (346 MHz) at 9.8 mT (5.9 mT).

\begin{figure*}[tbh]
	\centering
	\includegraphics[width=0.9\textwidth]{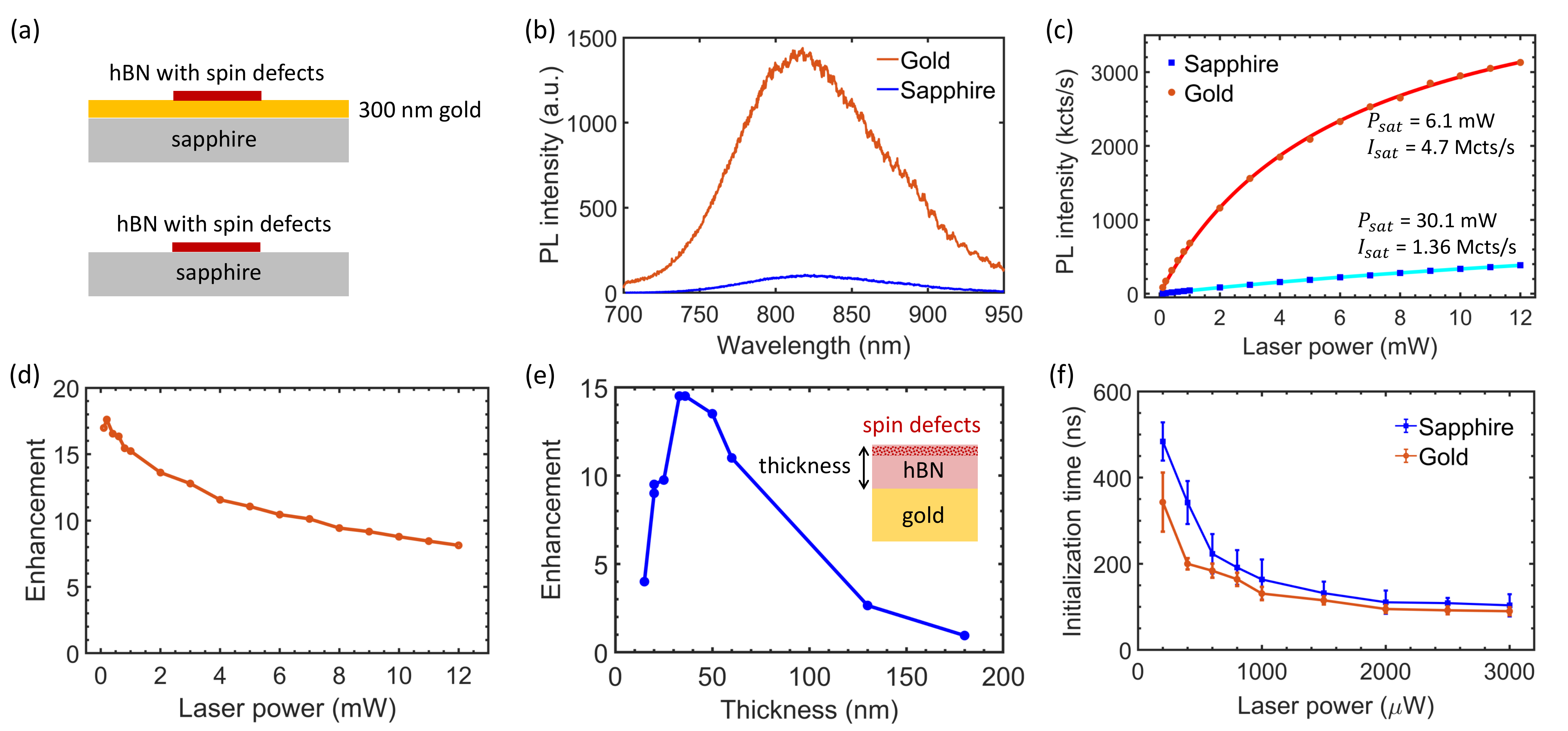}
	\caption{ (a) Illustrations of  hBN nanosheets with spin defects on top of  a 300-nm-thick gold film, and on top of a bare sapphire wafer for comparison.  (b) A comparison of the PL spectra of $V_B^-$ defects on a gold film and on a sapphire substrate. The gold film enhances the PL count rate substantially. The $V_B^-$ defects are generated by 2.5 keV He$^+$ ions. (c) PL intensities of $V_B^-$ defects on a gold film and a sapphire substrate as functions of the laser power. (d) PL enhancement at different laser powers. The enhancement is obtained by calculating the ratio of PL intensities the $V_B^-$ defects on the gold film and the sapphire substrate. The enhancement is up to 17 when the laser power is low. (e) Dependence of the PL enhancement on the thickness of the hBN nanosheet. The highest PL enhancement is obtained when the thickness is around 35 nm.  The $V_B^-$ defects are generated by 600 eV He$^+$ ions and are near the top surface of the hBN nanosheet. (f) The required time for optically polarizing  $V_B^-$ electron spins  as a function of the laser power.} \label{fig:PL}
\end{figure*}

It is highly desirable  to create spin defects as close to surface as possible without degrading the spin properties for nanoscale quantum sensing. This can  decrease the ultimate distance between a sample and the sensor, which can significantly improve the   signal. In Fig. \ref{fig:depth}, we study the formation of shallow $V_B^-$ defects by using He$^+$ ions. First, we use the Stopping and Range of Ions in Matter (SRIM) software to calculate the depths and densities of vacancies created with different ion energies from 200 eV to 3 keV (Fig. \ref{fig:depth} (a),(b)). The most probable depth is 3.5 nm, 6.4 nm, 15 nm, and 25 nm when the He$^+$ ion energy is 300 eV, 600 eV, 1.5 keV, and 2.5 keV, respectively. Then we perform the CW ODMR measurements on the samples with different doping depths. Here we use weak microwave driving to avoid the power broadening. So we can extract the nature linewidth of $V_B^-$ defects (Fig. \ref{fig:depth} (c)). All the ODMR spectra display similar linewidths as well as the contrasts, indicating  the hBN spin properties are nearly the same at different doping depths.

For sensing applications, the PL brightness  is an important factor that directly affects the sensitivity. Former theoretical studies indicate that the near-infrared optical transition of  $V_B^-$ defects is not an electric dipole-allowed transition and is hence relatively dark \cite{ivady2020ab,Reimers2020PhysRevB.102.144105}. In this context, improving their brightness is a crucial task. Here we utilize surface plasmons of a metallic film\cite{Shimizu2002PRLsurface,song2014photoluminescence,shalaginov2020chip} to enhance the brightness of the $V_B^-$. Surface plasmons are  collective oscillations of coupled eletromagnetic waves and free electrons on metallic surfaces. They have large localized electric and magnetic fields which can speed up both radiative and nonradiative decays.
We choose plasmonic enhancement because it can cover the whole broad PL spectral range of $V_B^-$ defects.  In addition, this method can utilize the metallic surface of our microwave waveguide and does not require complex nanofabrication. Our microwave waveguide is made of a 300-nm thick gold film prepared by electron-beam physical vapor deposition on top of a sapphire wafer. The width of the center microstrip is 50 $\mu$m. The hBN flakes with $V_B^-$ defects are transferred onto both the gold film and the sapphire substrate for comparison (Fig. \ref{fig:PL} (a)).   $V_B^-$ defects on both gold and sapphire surfaces display broad PL emission spectra around 810 nm (Fig. \ref{fig:PL} (b)). Remarkably, the $V_B^-$ defects on the gold film shows an order of magnitude higher PL intensity than those on the sapphire substrate under the same laser excitation. Fig. \ref{fig:PL} (c) and (d) present the PL intensities of these two samples and their PL ratio at different laser excitation powers. The experimental data in Fig. \ref{fig:PL} (c) is fit to $I=I_{sat}/(1+P_{sat}/P_{laser})$, where $I$ is the PL intensity of the $V_B^-$, $I_{sat}$ is the saturation PL intensity, $P_{laser}$ is the excitation laser power, and $P_{sat}$ is the saturation laser power. On a gold film, the $V_B^-$ defects show up to 17-fold enhancement of PL intensities under low-power excitation. We also observe a strong modification of the saturation behavior (Fig. \ref{fig:PL} (c)). The laser saturation power $P_{sat}$ is reduced by a factor of 5 and the saturation PL count rate $I_{sat}$ is increased by around 3.5 times.  These indicate that a gold film can improve the quantum efficiency of $V_B^-$ defects significantly.

\begin{figure}[tbh]
 	\centering
 	\includegraphics[width=0.5\textwidth]{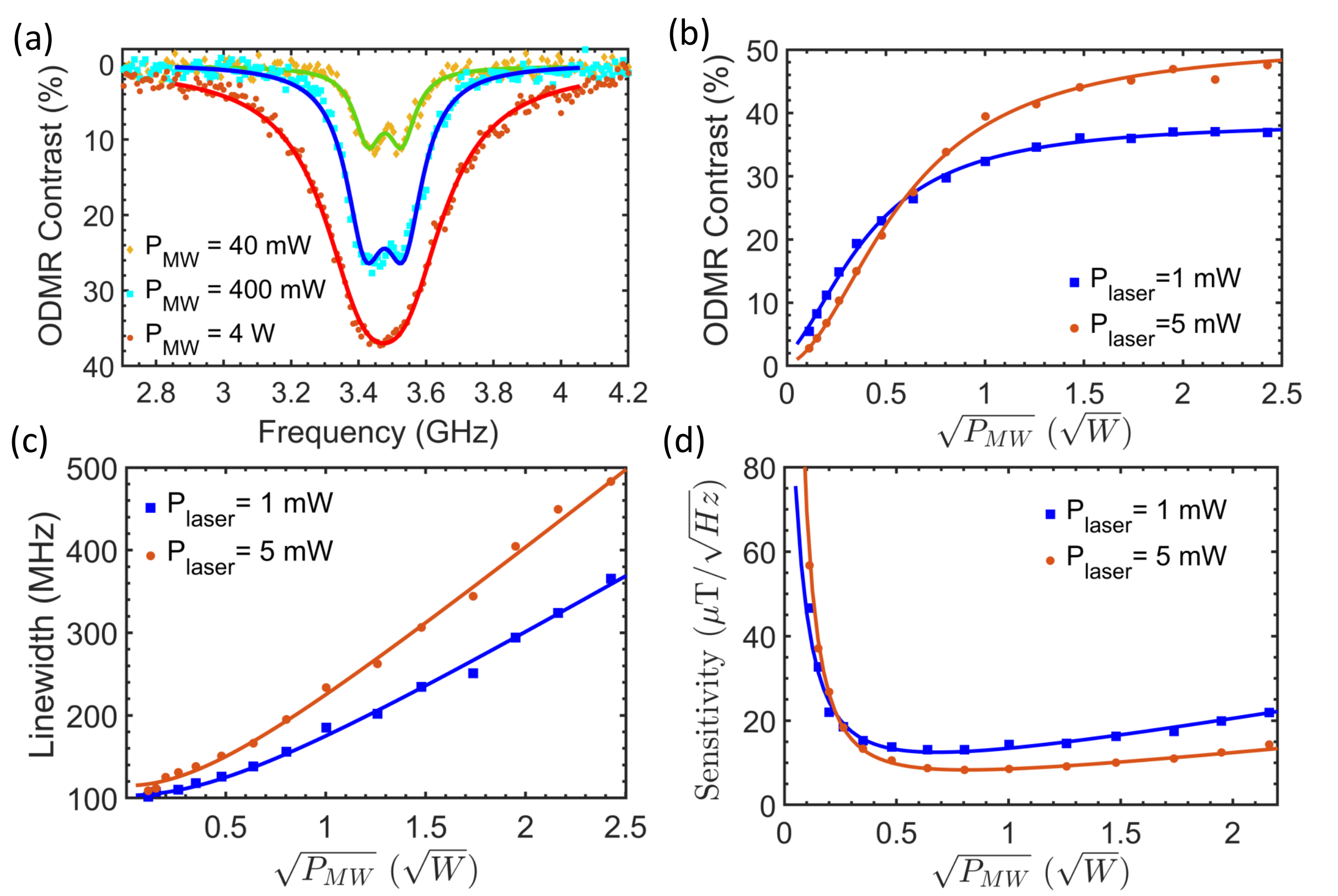}
 	\caption{ (a) Typical CW ODMR spectra of $V_B^-$ spin defects at different microwave powers. The laser excitation power is 1 mW. No external magnetic field is applied. (b) ODMR peak contrast as a function of the microwave power. (c) Microwave power dependence of the ODMR linewidth. (d) Magnetic field sensitivity as a function of the microwave power. The orange dots and blue squares are experimental data. Solid curves are fittings with theoretical models. } \label{fig:4}
\end{figure}

We also study the effect of the separation between the gold film and $V_B^-$ defects on the brightness enhancement. We transfer hBN flakes with different thickness onto the gold micostrip, so that we can get various distances between the $V_B^-$ defects and the gold film. The thickness of hBN nanosheets is measured by an atomic force microscope (AFM). The PL intensities are measured before and after transfer.  Here we use 600 eV He$^+$ ions to generate shallow $V_B^-$ with a most probable depth of 6.4 nm. So the average separation between the gold film and $V_B^-$ defects is equal to the thickness of hBN nanosheets subtracted by 6.4 nm.  As a result, we observe a strong thickness dependence of the PL enhancement. The highest enhancement is obtained when the hBN flake thickness is around 35 nm. Our result is consistent with the former result on plasmonic enhancement of quantum dots on a gold surface  \cite{song2014photoluminescence}. When a hBN nanosheet is too thin, the nonradiative decay dominates. There is also little brightness enhancement when the hBN nanosheet is too thick because the surface plasmonic modes decay exponentially away from the surface.  So there is an optimal thickness for plasmonic enhancement. In addition, we characterize the electron spin initialization time of $V_B^-$ defects on both gold and sapphire surfaces as a function of the power of the 532 nm excitation laser (see the ``Supporting Information'' for more details).  We find that the required spin initialization time is on the order of 100 ns. The gold film reduces the spin initialization time (Fig. \ref{fig:PL}(f)), which also speeds up quantum sensing.

Sensitivity is the most important parameter to determine the performance of a sensor. To measure an external static magnetic field with spin defects, a common way is to use CW ODMR to detect the Zeeman shifts of the spin sublevels caused by the magnetic field \cite{schirhagl2014nitrogen}. The precision to determine the magnetic field is directly affected by the photon count rate $R$, the ODMR contrast $C$ and the linewidth $\Delta\nu$, following the equation \cite{dreau2011avoiding}
\begin{equation}
\eta_B=A\times\frac{h}{g\mu_B}\times\frac{\Delta\nu}{C\sqrt{R}}, \label{sensitivity}
\end{equation}
where  $\mu_B$ is the Bohr magneton. In this expression, $A$ is a numerical parameter related to the specific lineshape function. For a Lorentzian profile, $A$ $\approx$ 0.77.  The values of $C$, $\Delta\nu$ and $R$ are further related to the microwave power and laser power \cite{dreau2011avoiding} (see ``Supporting Information''). To improve the detection sensitivity, it is crucial to increase the count rate and contrast as high as possible  without significant power broadening of the linewidth. 

Here we perform a group of CW ODMR measurements with various microwave powers and laser powers to find the optimal conditions for magnetic field sensing (Fig. \ref{fig:4}). Fig. \ref{fig:4} (a) presents three typical CW ODMR spectra at different microwave powers. Under low-power 40~mW microwave driving, we obtain an  ODMR contrast of $\sim$ 10$\%$. Such a low microwave power does not induce significant spectral power broadening. A natural linewidth can be extracted as $\sim$ 110 MHz. With an increasing microwave power, we first observe a significant improvement of the contrast without much spectral broadening. When we increase the microwave power further, the linewidth  broadening becomes severe. Fig. \ref{fig:4} (b)-(d) present the quantitative measurements of the ODMR contrast, linewidth and sensitivity as functions of the microwave power. The experimental results fit well with theoretical models as discussed in the Supporting Information. Here we perform the experiments at two different laser powers (1 mW and 5 mW). The 5 mW laser excitation gives a broader linewidth but a higher saturation ODMR contrast compared to those with 1 mW laser excitation. As a result, if we increase the microwave power, the sensitivity is first improved owing to the increase of the ODMR contrast and then becomes worse when the spectral power broadening dominates. The best sensitivity that we have achieved is about 8 $\mu \rm T/\sqrt{\rm Hz}$. This sensitivity is 10 times better than the former result with hBN spin defects (our system also has a better spatial resolution) \cite{gottscholl2021sub}.  This sensitivity  will be enough for studying many interesting phenomena in magnetic materials. For example, the magnetic field generated by a monolayer CrI$_3$ (a 2D van der Waals magnet) is on the order of 200 $\mu \rm T$ \cite{Thiel973}.

\begin{figure}[tbh]
	\centering
	\includegraphics[width=0.48\textwidth]{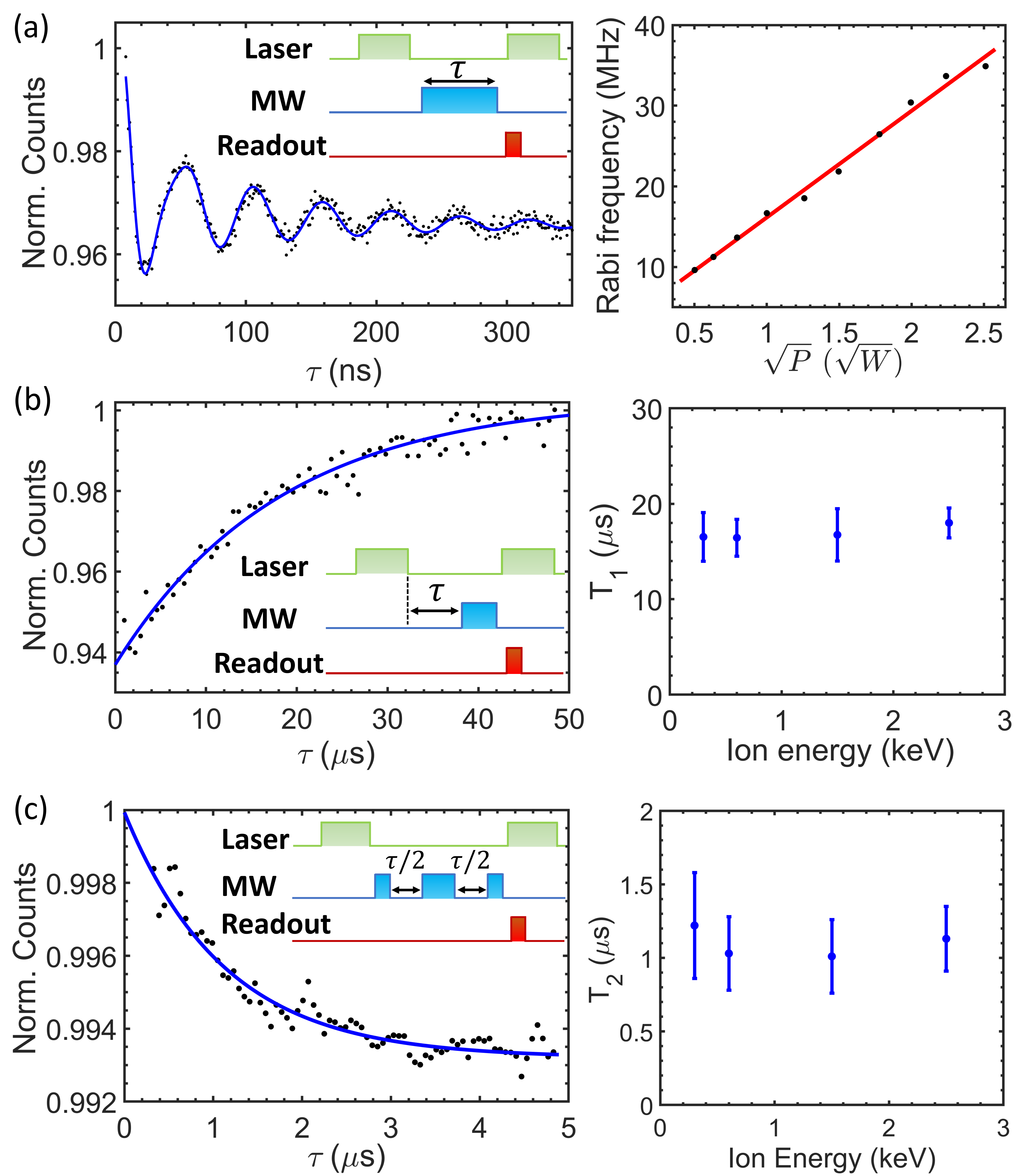}
	\caption{ Pulsed ODMR measurements of $V_B^-$ spin defects. A static magetic field of 13 mT perpendicular to the hBN nanosheet is applied to split  the  spin sublevels. (a) (left panel)  Rabi oscillation of $V_B^-$ spin defects.  (Right panel) Rabi frequency as a function of the microwave power.  (b) (Left panel) Measurement of the  spin-lattice relaxation time $T_1$. (right panel) $T_1$ of  $V_B^-$ defects with different depths. (c) (Left panel) Measurement of the spin-spin relaxation time $T_2$. (right panel) $T_2$ of the $V_B^-$ defects with different depths. }\label{fig:5}
\end{figure}

Finally, we perform  pulsed ODMR measurements to determine the spin-lattice relaxation time $T_1$ and the spin coherence time $T_2$ of the shallow $V_B^-$ defects generated by ion implantation. $T_1$ and $T_2$ of hBN spin defects have only been measured for neutron irradiated samples before \cite{gottscholl2021room,liu2021rabi}. It will be useful to know their values for our shallow spin defects created by ion implantation. In addition, pulsed ODMR measurements  are inevitable steps for realizing more complex sensing protocols. A pulsed ODMR measurement consists of optical initialization of the ground state, coherent manipulation of the spin state with microwave pulses, and optical readout of the final spin state. Here we add an external magnetic field of 13 mT to split two branches of the $V_B^-$ spin sublevels. m$_s$ = 1,0 states are used as the two-level spin system to carry out the spin coherent control. Fig. \ref{fig:5} (a) shows the Rabi oscillation as a function of the microwave power. The data is fit using $A+B_1 \exp(-\tau/T_1)\cos(2\pi f_1\tau+\phi_1)+B_2 \exp(-\tau/T_2) \cos(2\pi f_2\tau+\phi_2)$. We observe an oscillation with two Rabi frequencies, which are in the tens of megahertz range. The exponential decay of one oscillation component gives the spin-dephasing time of $T_2^*$ = 120 ns. To gain more insights of the spin properties of the $V_B^-$ defects at different depths, we measure the spin-lattice relaxation times $T_1$ and spin-spin relaxation times $T_2$ of the $V_B^-$ defects created with different ion implantation  energies. The pulse sequences are shown as insets in the left panels of  Fig. \ref{fig:5} (b),(c). By fitting the results, $T_1$ and $T_2$ are obtained as $\sim$ 17 $\mu s$ and $\sim$ 1.1 $\mu s$, respectively. As shown in Fig. \ref{fig:5} (b),(c), both $T_1$ and $T_2$ are independent of the ion energy, indicating the spin properties of the $V_B^-$ defects are nearly independent of the depth.

In conclusion, we have realized a significant improvement of the ODMR contrast and the brightness of  hBN $V_B^-$ spin defects. We observe a record-high ODMR contrast of 46$\%$, which is one order of magnitude  higher than the best former result with hBN. Our low-energy He$^+$ ion implanter can create very shallow $V_B^-$ defects close to the hBN nanosheet surface. Moreover, both CW and pulsed ODMR measurements display that their spin properties are nearly independent of the ion implantation energy. This result confirms the feasibility to create high-quality $V_B^-$ defects proximal to the hBN surface. In addition, we utilize the gold film surface plasmon to enhance the brightness of $V_B^-$ defects and obtain an up to 17-fold enhancement of the PL intensity. We also explore the effects of laser power and microwave power on the CW ODMR contrast and linewidth. With these, we achieve a CW ODMR sensitivity around 8 $\mu \rm T/\sqrt{\rm Hz}$. We expect that the PL can be enhanced further with plasmonic nanoantennas \cite{akselrod2014probing}, and the magnetic field sensitivity can be  improved by using more powerful pulsed sensing protocols. In addition, during the preparation of this manuscript, we became aware of a related work \cite{froch2021coupling} that reported a PL enhancement of hBN spin defects with a photonic cavity by a factor of $\sim 6$  at high NA collection and an ODMR contrast of about 5$\%$. The surface plasmon may be combined with a photonic cavity to obtain  better results in the future.
Our work strongly supports the promising potential of $V_B^-$ defects as a nanoscale sensor in a 2D material platform.


We thank supports from the Purdue Quantum Science and Engineering
Institute (PQSEI) seed grant and the DARPA QUEST program. We thank helpful discussions with Yuan Ping and Vladimir M. Shalaev.


%

\end{document}